# Drake's Rule as a Consequence of Approaching Channel Capacity

Alexey A. Shadrin[1,2], Dmitri V. Parkhomchuk[3*]
[1]Fachbereich Mathematik und Informatik, Freie Universität Berlin, Berlin, Germany
[2]Department of Vertebrate Genomics, Max Planck Institute for Molecular Genetics, Berlin, Germany
[3]Institut für Medizinische Genetik und Humangenetik der Charité – Universitätsmedizin Berlin, Berlin, Germany
Email: [*]pdmitri@hotmail.com

## Abstract

How mutations accumulate in genomes is the central question of molecular evolution theories. However, our understanding of this process is far from complete. Drake's rule is a notoriously universal property of genomes from microbes to mammals – the number of (functional) mutations per-genome per-generation is approximately constant within a phylum, despite orders of magnitude differences in genome sizes and diverse populations' properties. So far there is no concise explanation for this phenomenon. A formal model of storage of genetic information suggests that a genome of any species operates near its maximum informational storage capacity, and the mutation rate is near its upper limit, providing a simple explanation for the rule with minimal assumptions.

# Introduction

For about a hundred years the key parameter in modeling of Darwinian selection is "fitness" – it defines which organisms are left to live and reproduce in a population and which have to be eliminated. Alleles or mutations ("variants") are presumed to affect the fitness and a variant destiny (its frequency dynamics in a population) is traced with some mathematical models. There are numerous models with different assumptions about how to model real populations "correctly". For example the "Moran process" (Moran 1962) defines an elementary time step as either death or reproduction of a random individual – "overlapping generation", deriving the analytical solutions for some simple scenarios. Alternatively, the "Wright-Fisher model" (Durrett 2008) presumes the non-overlapping generations, such as annual plants. Then there are questions of how to calculate the cumulative fitness for a few independent variants, taking into account the effects of newly appearing variants, and many other subtleties. In the traditional models the fitness is "relative", without any fixed "baseline" – an individual cannot be assigned a fitness value ignoring the rest of the population. It is impossible to compare a fitness of an elephant to a fitness of yeast. Such fitness keeps no population history – a gain (or loss) of fitness for a whole population is untraceable, because after the gain the organisms are competing with each other in formally the same way. So the progressive evolution is presented as an opportunistic non-directional "Brownian" motion – fixation of accidental "positive" mutations. However, it would be tempting to have a measure, which is "absolute" – has a baseline and reflects the organismal complexity – the total "genetic information" or "evolutionary progress". On one hand, this measure would allow us to compare different species. However, what is more important, this measure would be a natural choice for the fitness function within a population for modeling.

Despite the numerousity of models their explanatory power remains arguably limited, so that in 1996 Ohta and Gillespie admitted "a looming crisis" – "all current theoretical models suffer either from assumptions that are not quite realistic or from an inability to account readily for all phenomena." (Ohta and Gillespie 1996). Probably the limits of the current models are rooted in the basic fitness definition, because if it is similar in all models, the reshuffling of other parameters will not drastically change the behavior and predictions on a fundamental level. Recently we proposed an information-theoretical model (Shadrin et al. 2013), which can provide a measure, which is "absolute", estimates the total genomic information and can be used for the fitness calculations, sensibly accounting for interactions of any number of variants in a genome. Such fitness function is the most essential difference of our model from the traditional approaches, while the modes of reproduction and other parameters are of secondary importance. Due to the novelty of such function we have to explore the model properties starting from the very basic considerations, omitting the moment phenomena, which are routinely considered in standard models, such as the influences of recombination, linkage, sexual selection, fluctuating environment, etc. Though clearly, such phenomena would be interesting to include in the subsequent development and to compare the results with traditional approaches.

Random mutations deteriorate the genomic information and must be compensated by selection. Here using simulation we evaluate some simple scenarios of such process under equilibrium condition. With some arguably plausible assumptions such process readily explains Drake's rule (Drake 1991; Drake et al. 1998; Sung et al. 2012). Here we address purely theoretical (postulated) phenomenon of Drake's rule, while its experimental validity for all species is quite different subject, which is not covered here. In fact the provided theory may suggest some clues about species, which are the "outliers" for the rule, having significant deviation from the trend.

## Methods

### Information in sequence patterns

The measure of genetic information (*GI*) was proposed by Schneider et al. (1986). It represents an adaptation of the entropy concept from the Shannon's information theory (IT) (Shannon 1948) to the context of biological sequences. During the last 25 years it became a popular tool for investigation of binding sites patterns (Schneider and Stephens 1990; Hertz and Stormo 1999).

The acceptable variability in each position (*P*) is defined by the frequencies of four nucleotides in an equilibrium population and quantified by Shannon's entropy:

$$H(P) = -\sum_{N \in \{A,G,C,T\}} f_N \log_2 f_N \qquad (1)$$

Where $f_B$, $B \in \{A, G, C, T\}$ is a frequency of nucleotide *B* at the position *P*. The genetic information for a single position is defined as: $GI(P) = 2 - H(P)$. One possible interpretation is that such function conveniently (additively and linearly) quantifies the amount of biases in equilibrium distribution of alleles.

For technical simplicity, we (after Schneider et al.) assume independent positions in patterns, otherwise we would have to deal with general "typical sets" and the *GI* computation would be more complicated. However, there are no indications that assuming some positional dependencies in patterns would drastically influence the main conclusions. While covariable sites are known, significantly correlated sites can be grouped in "pseudo-sites" (now with more than four states) so that correlations can be canceled (concealed) with a proper basis selection, so the generality of our conclusions here is beyond the influences of such covariable sites. Recently we showed (Shadrin et al. 2013) that the sum of *GI*s can serve as a measure of positional information. This "additivity" should not be confused with a simple additivity of Shannon's entropy – the problem is to prove that the sum of *GI*s for a functional site (or a genome) is linearly linked to the "positional information" (specificity of molecular interactions). One could use some other measure of frequencies biases – why is the defined one "fundamental"? In order for the sum of site's *GI*s to have the informational meaning, the number of possible functional sequences for the site (the size of its "typical set") must depend exponentially on the defined site's variability (the value reciprocal to the sum of *GI*s). This exponential dependence is the non-trivial result of the IT (Shannon-McMillan-Breiman theorem). The corresponding "natural choice" of the logarithmic function for information measure is discussed in details in the classical Shannon's paper (Shannon 1948). With such well-defined positional information measure it is possible to build a formal ("mechanistic") model of "molecular machines" evolution.

Then we can use the position-specific *GI*s to calculate the total amount of information contained in a genome as a simple sum over all positions:

$$GI_{total} = 2L + \sum_{j=1}^{L} \sum_{B \in \{A,G,C,T\}} f_{jB} \log_2 f_{jB} \qquad (2)$$

Where $f_{jB}$, $1 \leq j \leq L$, $B \in \{A, G, C, T\}$ is the frequency of nucleotide *B* at the position *j*. Let's also define an average density of genetic information in a population as $GI_\rho = GI_{total}/L$ bits per site. It is obvious that $0 \leq GI_\rho \leq 2$.

Hence a functional site (or a genome) is represented by the corresponding pattern – "*GI* profile", so that *GI*s and the corresponding 4-vectors of the acceptable equilibrium frequencies are defined in each position. As we discussed in (Shadrin et al. 2013) the equilibrium condition is important for the correct *GI* definition and measurement, while, in general, real populations are far from the equilibrium. Importantly, the *GI* profile is the "prior", inherent property of molecular functionality, for example a protein domain can be functional only within a certain set of sequences, e.g. in *GI* terms, a conserved domain has high *GI* value and a small "typical set". So we posit that a given species is fully characterized by a set of all possible sequences, which produce the species-specific phenotype – defining the typical set. It is clear that this set is much smaller than all possible random sequences. However, in general, it is much larger than a realistic population size. For this reason we need to simulate the equilibrium – to enumerate the complete typical set. Then the average density $GI_\rho$ cannot be significantly different in close species – functional genes are conserved similarly, unless some novel mechanisms of molecular functioning are introduced. The actual variability in a population depends on this predefined *GI* and a population history. The equilibrium population, which we simulated here (effectively canceling out, "erasing" the history influence, revealing the unobscured "pure functionality" profile), is necessary for the correct *GI* measurement (the knowledge of complete typical set) and the determination of an "error threshold". However, a slice (small subset) of such population will have the same mutational properties as the whole equilibrium population, but smaller variability. Such subset represents a realistic population, which recently (relatively to mutation rate) underwent a bottleneck and experienced a "founder effect" – all individuals are closely related through a few recent population founders. The variability in this subset does not reflect correctly the *GI* profile. However, this profile still "exists", though more in a platonic sense. It could be revealed if this subset was allowed to diverge for sufficiently long time without any disruptive events. This equilibrium population shows the principle limit on the maintainable pattern (revealing the full typical set, total *GI*, quantifying the total amount of biases), which is then defined solely by the mutation rate and reproduction/selection properties of the population, since the dynamical part ("history") is excluded. It is clear that, with other things being equal, this limit plays the same limiting role for the "collapsed" population (after a bottleneck). We can imagine that under the influence of mutagenesis a realistic population is drifting inside a large typical set. Nonetheless, this is much more restrained drift in comparison with the drifting in a space of all possible sequences by random walk – as the neutral theory suggests (for the majority of accumulating mutations). However, since a typical set can be huge, in general, the drift in it might give an impression of a random drift.

Such modes of mutagenesis and maintenance of variability are similar to those in quasispecies theories: "The quasispecies concept becomes important whenever mutation rates are high. This is often the case in viral and bacterial populations." (Nowak 1992). In these theories a population is represented by a "cloud" of diverged genotypes. However, the distinction between "normal" species and quasispecies is blurred, and nothing can prevent us from viewing a "normal" population as the aforementioned subset of quasispecies (in the process of divergence). Here we assume that this mode of high mutation rate is precisely the one, which deserves careful examination in the large genomes of higher organisms as well – what matters is the mutation rate per-genome per-generation, and as we now know, this parameter is quite large in mammals also – about few hundreds mutations, with few in coding regions – actually that is the main point of Drake's rule. For simplicity we presume, that the selection has an opportunity to act compensatory (to increase *GI*) only in-between generations, ignoring possible germ-line selection issues. That is the reason for focusing on the per-genome per-generation mutation rates – the selection does not "see" a genome size or per-base mutation rate. What it does "see" is the effect of a number of functional mutations, which it tries to compensate through genetic deaths - removal of the genomes from a population. So the natural "units" for selection actions are a genome and a bunch of mutations in it. In comparison, the quasispecies theory is used to address the evolution of HIV with 1-10 mutations per division, so from the selection point of view the functional impact (at least in *GI* terms) is comparable. As pointed out by Nowak (1992), HIV population "seems to operate very close to its error threshold". The existence of this "threshold" is our main postulate here. However, we apply it to all species, and with the provided IT framework, such threshold seems to be well-defined and ready for modeling. The main difference between virus and mammals populations seems to be the generation time and the genome size – the virus genotype "cloud" can be readily observed empirically. However, to generate the actual equilibrium "cloud" for a large, slowly replicating genome would take the astronomically large time and population size – equivalent to enumerating the full typical set. Nevertheless, this does not mean that we cannot explore the properties of this limit theoretically and then assume that these properties are applicable to the aforementioned population slice. The equilibrium mode of maintaining variability is considered in quasispecies theories too, and after we introduce the pattern definition and the measure of genetic information with fitness function we arrive to our model. However, in the quasispecies theory the fitness is defined for the whole population of mutants, not for individuals (Nowak 1992).

In our model we assume that (for $GI_\rho < 2$ bit) a large number of allowable sequences (constituting a typical set) are nearly "synonymous" and can coexist in a population in the case of the equilibrium maintenance evolution. However, they are not completely synonymous so that the selection can maintain a pattern by discarding the most deviant ("atypical") sequences. The advantage of the model is that it allows meaningful evaluation of the information contained in a pattern (or a genome). Furthermore, given the defined weight matrices of a desired conservation profile, the model provides selective values of individuals considering all mutations, present and *de novo*. We showed (Shadrin et al. 2013) that the substitution rate in functional sequences can be arbitrary close to the neutral rate and the fraction of positive mutation can be high in general. About 50% of the retained mutations must be "positive" - a trivial requirement for the balance of *GI*.

How realistic is such selection and fitness modeling? It is as realistic as Turing machine. This model, analogously to studying of evolution with Turing machines (Feverati and Musso 2008), can be described as a population of machines operating on symbol sequences (of limited length), reading out positional information recognizing corresponding patterns (via typical sets, technically, for a general typical set the assumption of positional independence is not necessary) of molecular interactions and calculating high-level phenotype. However, it seems that our machine is closer to describing the "molecular computations" through patterns recognition in comparison with the sequential algorithmic Turing machine. For the purposes of this investigation we don't have to specify the phenotype calculations per se – once we define the patterns and typical sets in a "genome" we can address the problem of their maintenance or evolution (e.g. the cost or speed of patterns preservation or change). Here we focus on the maintenance properties, treating such machines as genetic information storage devices, which must resist random noise of mutagenesis. The only computation is done for selection actions – genome "typicality" is used as fitness, accounting for all variants in a genome (Equation 3). As could be expected our fitness function is similar to the traditional one in its basic "common sense" features – for example a mutation in highly conserved site (high *GI*) will drop the fitness significantly. Notably, in this model all sites and variants are functional – there is no need to postulate "neutral" (Kimura 1983) or "near-neutral" (Ohta 1973) variants (to explain the high rates of sequence evolution) – in our case, the equilibrium can be interpreted as the cumulative neutrality of all mutations (remained in a population), while assuming the individual neutrality of all or most mutations would be throwing the baby out with the bathwater.

### Simulation terms

An organism in the simulation is represented by the nucleotide sequence of given length (*L*), $O = [B_1, B_2, ..., B_L]$, where $\forall\ i \in [1,L]$, $B_i \in \{A,G,C,T\}$. A population is a set of organisms of the same length.

Parameters, which govern the process of simulation, are shown in Table 1. The mutational bias is included in the code for universality, but has no effect on the trends we investigate here.

Each organism ($O$) in a population can be associated with a weight specified by the weight matrix $W$:

$$W(O) = W([B_1, B_2, \ldots, B_L]) = \sum_{i=1}^{L} W_i(B_i) \quad (3)$$

A "typical" probability is the expected probability of a sequence for a given *GI*-profile. It can be calculated through multiplication of the expected frequencies for corresponding positions. Here, for computational convenience, we define the fitness as a sum of position-specific weights, which, for our purposes, is equivalent to the multiplication if we had used logarithms of frequencies. Technically, allele-specific fitness contributions can be additive, multiplicative or any other formulations; such (potentially interesting) complications can influence only the shape of the resulting *GI*-profile and its stability (fluctuations); they do not affect the existence of the mean density and the independence of the population size. However, for example the specifics of reproductive success dependencies obviously can be important for the dynamical part – before the equilibrium is reached.

We do not know the resulting *GI*-profile before the simulation is performed. Hence the weight matrix defines a general direction of pattern conservation by selection, not the actual *GI*-profile per se.

**Table 1. Simulation parameters.**

| Notation | Description |
|---|---|
| $N$ | Number of organisms in the population (population size). |
| $L$ | Length (number of bases) of genome of each organism in the population. |
| $n_d$ | Number of descendants each organism produces in a single round of reproduction. |
| $P_m$ | Probability of mutation per base. |
| $P_{ti}$ | Probability that an occurred mutation will be a transition mutation. |
| $W = (W_j \mid j \in [1,L])$ | Selection weights of nucleotides in each position. Where $W_j = (w_{jA}, w_{jG}, w_{jC}, w_{jT})$, $W_j(B) = w_{jB}$, $B \in \{A,G,C,T\}$ – selection weight of the corresponding nucleotide $B$ in $j$-th position. |

This weight is used to determine preferences of selection, which tries to maintain a pattern. In our experience, the particular recipes for selection actions (e.g. probabilistic/deterministic) and reproduction modes (overlapping/non-overlapping generations) play little role for the described trends, as long as the main purpose of these actions is to maintain a pattern – a biased frequencies distribution, while the opposing force – random mutagenesis tries to flatten the bias. Each mutations round decreases the genomes "typicality", in average. So a more "typical" genome has higher reproductive success, because its progeny is more likely to stay typical and avoid elimination. As we mentioned *GI* can be viewed as a convenient measure of functionally acceptable variants frequencies biases. Such fitness definition, in our opinion, is the key departure from traditional models. For example it seems to be inherently difficult to approach Drake's rule explanation with a fitness function which is relative – it has no information on organisms' degree of complexity, hence, taken alone, it is "blind" to a genome size. In our case the total *GI* – organismal complexity is measured by the amount of pattern's (functionally acceptable) biases. It seems to be intuitively appealing quantification – the larger the total amount of biases (further from the flat distribution) – the higher the information content and more it takes to maintain it. However, such approach is a necessary simplification – it works under the assumption that the rest ("higher order") information unfolding processes are approximately the same, which should work, at least for similar species.

Presumably, the sophisticated error corrections mechanisms such as DNA repair constitute a biological burden. So we ask: what is the maximum mutation rate, which is compatible with a given total *GI*? The differences of $GI_\rho$ of functional sequences are assumed to be small for close species. Formally, for our phenotype-calculating machines, the conservation of *GI* is equivalent to the whole phenotype conservation, because as we reasoned in (Shadrin et al. 2013) the *GI* conservation preserves positional information of molecular interactions, so that a phenotype is mechanistically derived from the whole genome pattern.

### Simulation Process

The entire simulation process can be divided into three successive stages: initialization, spawning and selection. The initialization occurs only in the very beginning and then the spawning and selection are repeated in a loop until the simulation process is stopped.

*Initialization:* the initial population, consisting of *N* organisms of length *L* is generated. All organisms in the initial population are identical and have maximum possible weight according to matrix *W*, i.e. at each position *j* of each organism stands a nucleotide $B_j$: $B_j = \left[ B \mid w_{jB} = \max_B \left( w_{jB}, B \in \{A, G, C, T\} \right) \right]$.

*Spawning:* the progeny is spawned. Each organism in the population produces $n_d$ descendants (here we consider in detail only the case of binary fission, i.e. when $n_d = 2$). A descendant organism has the same length as its parent and is obtained by the copying of the parental sequence with a certain probability of mutation ($P_m$) and with a bias of mutational spectrum ($P_{ti}$). The parental organism is excluded from the population after the reproduction, so the generations are non-overlapping and after this step the population consists of $n_d N$ organisms.

*Selection:* the selection reduces the number of organisms in the population back to the initial size. It acts deterministically, leaving *N* organisms, whose weight *W(O)* is larger.

The choice of procedure of the initial population generation does not affect the steady state of the simulation process, so we can simply generate a random initial population. However, generating the initial population as described above will provide the faster convergence to the steady state – the equilibrium condition, which reveals the "error threshold" – the goal of our experiments. The above mode of reproduction describes the non-overlapping generations, for the simplicity of defining and counting mutations; however, we experimented with other regimes, including the overlapping generations similarly to Moran model, and found the trends invariant.

## Results

### GI Behavior in the Course of Simulation

Immediately after the initialization stage of the simulation the $GI_\rho$ of the population according to formula (2) is equal to 2 bits, because all organisms are identical. However, as we discussed earlier that is not the "correct" functional *GI*, but a formally computed value in the course of simulation. If we start the simulation process as described above with the probability of mutation $P_m$ high enough to allow occurring mutations to propagate in the population, then the diversity will emerge and $GI_\rho$ will start to decrease. While reducing, $GI_\rho$ will finally reach the level, when the mutagenesis is balanced by the force of selection, and in consequent iterations will fluctuate in a vicinity of some value. The existence of the balance (mean $GI_\rho$) is clear because the capacity (the averaged effect) of random mutagenesis to decrease *GI* monotonically drops from some value at $GI_\rho = 2$, to zero at $GI_\rho = 0$, while the corresponding selection capacity to increase *GI* behaves reciprocally – having non-zero value at $GI_\rho = 0$ and zero at $GI_\rho = 2$. Thus these two functions intersect at some equilibrium point. In our numerical experiments we consider that the population is already in the equilibrium state if during the last T generations (T = 100 in our tests) two conditions are met: the sum of all $GI_\rho$ changes between consequent generations is less than a specified threshold (1e-3 in our tests), the maximum number of consequent generations increasing/decreasing $GI_\rho$ is less than 0.1*T.

Observable magnitude of $GI_\rho$ fluctuations around the equilibrium value depends on the size of the population, but the equilibrium value per se does not depend on it, which is natural to expect for the population maintaining constant variant frequencies. So setting the size of the population (*N*) large enough we can identify the moment of equilibration and equilibrium value of $GI_\rho$ with the required precision. Even if we assume more complicated scenario where the fluctuations are not settling down, the aforementioned capacities of mutagenesis and selection to change *GI* cannot depend significantly on the population size. They operate on the variants frequencies, which are disentangled from the absolute population size, hence the balance (even if it is the dynamic balance) between these two forces is also free from the population size dependence. We will call the state of the simulation when the population has already reached equilibrium as *GI*-steady state and denote the mean value of $GI_\rho$ in equilibrium population as $GI_{steady}$. A biological interpretation is that it is a given species maintainable *GI* value. It can be called a "mutation-selection balance", however, it is clearly different from Fisher's balance (Crow 1986), who considered a single site – in our case the balance is due to the compensatory effects of multiple positive and negative mutations. It should be, however, clearly understood, that the word "steady" here concerns only the total genetic information (and hence the phenotype), the genomes remain variable, because new mutations still appear with the steady rate. The "molecular clock" is ticking – and its empirical steadiness on the evolutionary scale is another indirect hint that the average *GI* density is a slowly varying parameter. For example, mutations are more frequent in a position with the lower *GI* value, so if the density would fluctuate strongly on the evolutionary scale, the clock would behave erratically. As we argued (Shadrin et al. 2013), *GI* increasing (positive) mutations constitute a significant fraction of random mutations (especially when *GI* in a position is low), thus allowing the same fraction (in the *GI* equivalent) of negative mutations to remain in the population. The monotonous molecular clock is a simple prediction of the provided model. Alternatively it can be explained by the neutrality assumption, which seems to be an oversimplification of reality. Also the provided model shows that the steadiness of the clock is intimately connected with Drake's rule and the "error threshold", while the neutral theory is inherently unable to make such connections. The convergence of $GI_\rho$ for different parameters is presented in Figure 1.

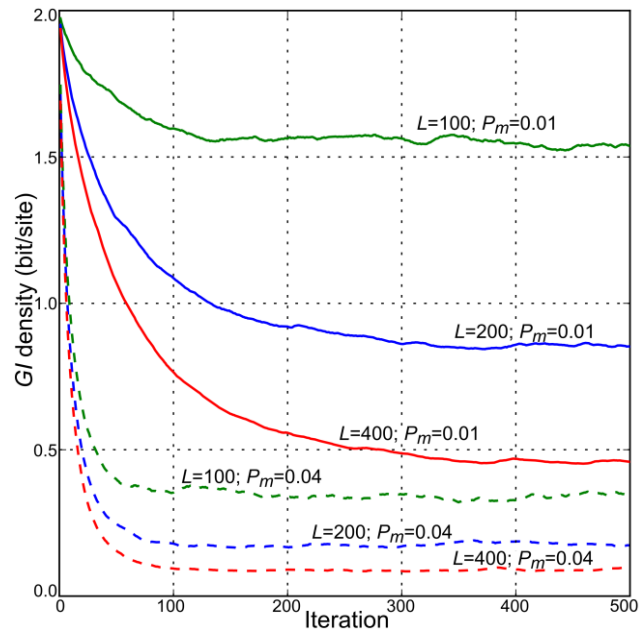

**Figure 1. Convergence of $GI_\rho$ for different parameters.**
Common parameters for all demonstrated cases are: $N = 1000$; $n_d = 2$; $P_{ti} = 2/3$; $W = (W_j = (0.8, 0.2, 0, 0)$ if $j$ is even, else $W_j = (0.5, 0.3, 0.1, 0.1))$. Color determines organism length (*L*): green corresponds to $L = 100$, blue to $L = 200$ and red to $L = 400$. Line style determines probability of mutation per base ($P_m$): solid – $P_m = 0.01$, dashed – $P_m = 0.04$.

## Counting Mutations

In the simulation the number of fixated mutations, i.e. the observed mutations per generation can be counted directly. Following the common notation we denote the number of mutations per generation per base as $u_b$ and the mutation rate per generation per genome as $u_g$. Despite the fact that the values $u_b$ and $P_m$ are closely related, $u_b$ is always less or equal than $P_m$, since the organisms with more mutations are more likely to be eliminated at the selection stage.

Now let's look at somewhat inverse experiment: we can fix the value of $GI_{steady}$ and all parameters from Table 1 except $P_m$, and then numerically find the value of $P_m$ which corresponds to the fixed parameters. This procedure can be performed for different lengths of organism (*L*) while maintaining the same values of all other parameters ($N$, $n_d$, $P_{ti}$, $W$, $GI_{steady}$). The fixed parameters in our experiment were set to: $N = 1000$; $n_d = 2$; $P_{ti} = 2/3$; $W = (W_j = (0.8, 0.2, 0, 0)$ if $j$ is even, else $W_j = (0.5, 0.3, 0.1, 0.1)$); $GI_{steady} = 1.6$. The experiment was performed for *L* values 64, 128, 256, 512 and 1024. Then we estimated the number of mutations observed in the *GI*-steady state and compared $u_b$ and $u_g$ parameters for different genome lengths. The results are summarized in Figure 2.

We also found the dependence of $GI_{total}$ vs. $GI_\rho$ when the mutations rate ($P_m$) is fixed while the genome size (*L*) varies (Figure 3). We considered the populations with genome sizes from 100 to 1100 with the step of 20. All other parameters of these populations were identical, namely: $N = 1000$; $n_d = 2$; $P_{ti} = 2/3$; $W = (W_j=(0.8, 0.2, 0, 0)$ if $j$ is even, else $W_j=(0.5, 0.3, 0.1, 0.1))$ and mutation rate $P_m$ was set to 0.007. For each population its *GI* density (*GI*$_\rho$) and the total *GI* (*GI*$_{total}$) were averaged over 1000 generations after the population reached *GI*-steady state.

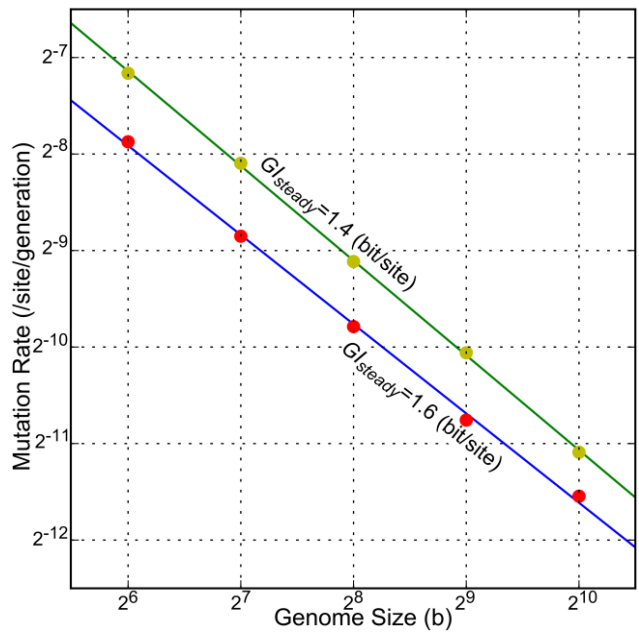

**Figure 2. Relationship between the mutation rate per site per generation ($u_b$) and the genome size (*L*) observed in the simulation.**
Red points – $GI_{steady} = 1.6$ bit per site, blue line – the corresponding regression line, based on the equation: $\log_2 u_b = -2.337 - 0.929 \log_2 L$ ($r^2 = 0.9993$).
Yellow points – $GI_{steady} = 1.4$ bit per site, green line – the corresponding regression, based on the equation: $\log_2 u_b = -1.251 - 0.982 \log_2 L$ ($r^2 = 0.9998$).

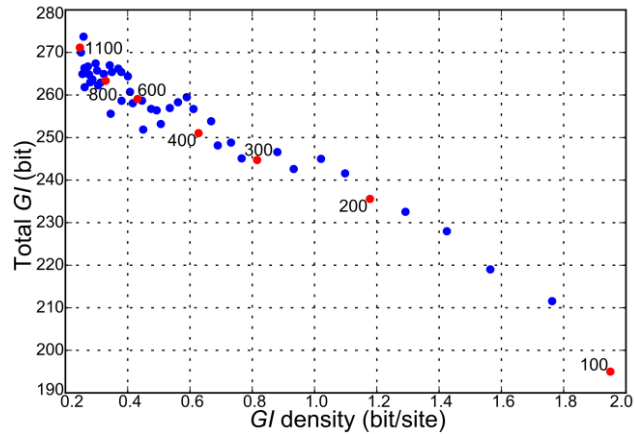

**Figure 3. Dependence of total *GI* (*GI*$_{total}$) on equilibrium *GI* density (*GI*$_\rho$) for a fixed mutation rate and different genome sizes (*L*).**
Each point represents a population with organisms having genome of size $L \in$ [100, 120, …, 1080, 1100]. For convenience of orientation some points are colored in red and the genome size of corresponding population is labeled.

Through defining different weights matrices (Equation 3) we tested different scenarios of the density distribution: with homogeneous *GI* distribution in a genome, and bimodal – one half of a genome consisting of highly conserved ("lethal") sites – to model the regions such as conserved protein domains and the other half consisting of weakly conserved sites, to model the variable parts of proteins and weakly conserved non-coding regulatory DNA. The meaning of the obtained results for the average *GI* density versus mutation rates is the same, so that the actual distribution does not affect the described trends.

## Discussion and Conclusion

Naturally the model's applicability to the evolution of real molecular machines should be thoroughly investigated since any formal model has its limits. For example particular alleles might interact in more complex ways than described by the model, though, as we mentioned, such interaction can still be accounted for, by constructing more complex typical set, without simplifying assumptions we used. However, these complications cannot influence the basic conclusions. And since the model provides simple explanations for observable phenomena, with the minimal number of parameters and assumptions, and, in principle, is realizable in a hardware, which operates similarly to our understanding of molecular interactions, we believe that it fairly captures the general properties of real genetic systems. Interestingly, the model can be considered as a simple generalization of Hardy–Weinberg equilibrium (HWE) (Hardy 2003), explicitly including functional sites and their maintenance selection. This may explain the persistent (about half-century) illusion of the neutrality – in usual tests (e.g. Tajima's D (Tajima 1989)), the mutations in an equilibrium population will "pretend" to be neutral, so such criterions actually test for the (local – in the case of recombining population) equilibrium condition, rather than for the individual mutations neutrality. Below we discuss some possible consequences for our understanding of real genetic systems, assuming that the model is sufficiently valid.

To prevent unnecessary criticisms, we have to admit that we are discussing some possible features of an idealized population described by the model; the correspondence to real genetic systems reflects the practical value of the model (such assessment is often influenced by subjective personal "tastes"), which is of secondary importance. For example the "heated" debates of neutralists vs. selectionists – it seems that our view reconciles both camps – the evolution is mostly neutral (in stasis), though this neutrality is maintained by the selection of the most typical individuals. Technically, it is a normal epistemological practice to explore abstract models regardless of their immediate relevance to "reality", since the corresponding applicability domains can be rather specific and not yet well established and delimited. For example the "strong selection", which leads to "selective sweep", is a non-equilibrium event and it is out of the equilibrium model applicability domain. In terms of *GI*, such event alone provides only 2 bits of *GI* for a given site, for the price of the total population replacement (roughly speaking). Such events are equivalent to considering a changing environment. Under the model's assumptions of the constant environment and the infinite time equilibrium, all such events would occur and be settled down. However, even for the changing environment situation, we propose that the model describes the "background" of such (presumably relatively rare) events. The number of such events must be limited by Haldane-type arguments (Haldane 1957), so we assume that the rest of mutational background can be better described by the provided model, than by the neutral approximation. In fact, according to the model a mutation per se, with any selective value, while changing individual organism's typicality or fitness, cannot increase the amount of total *GI* in the equilibrium population – the phenomenon we explain below. Hence, in this work, the model has well defined restricted applicability domain. However it is straightforward to extend it to certain non-equilibrium scenarios such as the abrupt or gradual changes of *GI*-profile, simulating changing environment. Admittedly most of variants in real populations are of weak effect, however, their number can be quite large, so that their collective effects can be far from negligible (as in the neutral theory) and our theory suggests a consistent way of accounting for such effects.

According to Drake (Drake et al. 1998) the genomic mutation rate "is likely to be determined by deep general forces, perhaps by a balance between the usually deleterious effects of mutation and the physiological costs of further reducing mutation rates". As can be seen, Drake correctly did not include considerations for adaptive properties of evolution, practically solving the problem, hinting that it is rather the maintenance-related phenomenon, and once we interpret the maintenance as the equilibrium in alleles frequencies – the main property of our model – the population size is obviously out of the equation (as in the case of HWE).

The key assumption for Drake's rule explanation is that the total genomic information is saturated to its maximum maintainable value, or reciprocally and equivalently, the mutation rate is near its upper limit for a given total *GI* of species. The mutation rates and thus the total *GI* are assumed to change slowly on evolutionary time scale. We hypothesize that the rate decrease is a basic event required for progressive evolution, and it is promptly followed by the gain in total *GI*, restoring the equilibrium. The equilibrium can be regained "quickly" (~100 generations, judging by the speed of convergence to the steady state in Figure 1). One difficult question is how to motivate the stability of the mutation rate for a given species. For the rate decrease, we can assume that it might happen due to the large difference of the time scale of two phenomena. The first is a merely long-term advantage of the lowered mutation rate – some generations must pass to fill newly accessible *GI* (if a niche requires it, which does not have to be the case, in general). The second is the immediate disadvantage – "physiological costs" – since the lower rate, in principle, must be associated with a slower replication rate and/or additional energy expenditures. On the other hand, why the rates are not degrading, if increasing the rate might bring a fast advantage, but only a long-term disadvantage? At this point we can only speculate that for higher organisms, an increased somatic mutagenesis might cause also the short-term disadvantage, preventing the rate degradation (e.g. somatic mutations theories of aging or carcinogenesis). Beside the somatic mutagenesis, we could imagine any other selectively important phenotype, which is somehow linked to the changes in mutation rate. The other idea is that while the rate decrease must come at some "physiological costs", the way back is not that easy – a mutation, which degrades the rate will not necessarily reduce the "physiological costs" back to the previous values. Such mutation must be rather specific "back-mutation" or, more likely, a number of them, making it improbable to achieve both – the rate increase and the corresponding costs reduction. Hence the rates can only go down, locked from above by both short- and long-term disadvantages. Alternatively, the rate maintenance might require a regular population renewal, described below. Naturally there are examples of regressive evolution, which are not "interesting" for us here – such evolution can be easily caused, for example, by moving to a simpler niche (habitat) – "use it or lose it".

Hypothetically a change of mutation rate would remold a species phenotype (suggesting an explanation for "punctuated equilibrium" phenomenon), since small relative change of the mutation rate can provide significant absolute change in total accessible *GI* and correspondingly a significant change of the phenotype. In principle it is assessable experimentally – if we were able to select flies (for example) for the lowered mutation rate, then the model predicts that such population has the capacity to produce more advanced species of flies. The problem is, however, that the population must be challenged with the proper external conditions, which could cause the evolutionary progress, promoting an increase of complexity.

It is natural to expect that the rates occupy discreet values, due to the discreet nature of corresponding modifying mutations and their (presumably) limited number. We could also hypothesize about speciation scenarios: suppose that in a large population the rates are heterogeneous and mixed, so that the population has some average rate. Then after a "founder" splits off, he produces a new population, which can have the rate different from the main population, leading to the fast phenotype changes.

The evolution of the (functional) genome size is presumed to occur through gene duplications (Ohno 1974), so that "gene families" grow in size. That also motivates our postulate of slow changes of $GI_\rho$ for functional sequences – new sequences perform molecular functions similar to the original. The provided theory readily predicts that whole or partial genome duplications would lead to an increased rate of sequence evolution and to a subsequent shrinking back of the (functional) genome size, loosing extra gene copies, due to inability to maintain higher total *GI* without changes of the mutation rates. The evolutionary progress (an increase in complexity and *GI*) is happening not due to duplications per se (which are relatively frequent events), but due to the mutation rate decrease and/or the adoption of the lower $GI_\rho$ functionality (Figure 3) – these changes are assumed to be "slow". However, the progressive evolution is naturally intervened with the external conditions (niche or habitat), and it must be sufficiently complex to support the increase of species complexity. Duplications might also cause a reproductive isolation, hence, together with the founder-specific mutation rate hypothesis this might be a path to speciation and progressive evolution (when the founder retains the lowered mutation rate).

We suggest that the "channel capacity" notion of IT is sufficiently deep and general principle to provide the desired understanding of Drake's rule. The notion also allows quantitative modeling of the process. Channel capacity is the upper bound for the information transmission rate for a given noise level. Practical solutions for information transmission are somewhat below this theoretical limit, and considerable engineering efforts are dedicated to approach the limit, simply because being closer to the limit saves energy. Hence another basic consideration is that if the nature would not use the genomic informational capacity to its full extent, it would not be "thrifty" – why waste resources on the unused capacity – the thriftiness should be favored by selection (though there are some opposing ideas of "selfish" or "parasitic" sequences). If we presume that the early genetic systems operated at the "error threshold", it is not clear at which moment and for what reasons this threshold was abandoned. It seems to be the most thrifty and the fastest way to progress - to stay always on the threshold, which is moving up due to the enhancements of replication fidelity and possibly other mechanisms. In fact, considering the "costs", it is difficult to come up even with an artificial reason to push the fidelity beyond necessity. Thus, unless we discover some good motives for this reason, we have to admit (following Occam's lines) that the contemporary species are also at the "error threshold". It seems that ignoring this fundamental threshold would make the evolutionary modeling critically incomplete.

Intriguingly, in IT the problem of approaching the channel capacity limit has no general solution applicable to all practical situations, as it relates to the problem of achieving best compression rate, and, in practice, is limited by the memory and computational costs. That creates a recursive, self-referenced evolutionary system: in order to become more effective, resources must be invested in some analogs of memory and computations, and these resources, in turn, must be used in the thriftiest way, i.e. optimizing the optimization, etc., producing a Gödel-like system. Analogously to Chaitin's proposals (Chaitin 2012), we can speculate that molecular machines have an infinite field for exercising the mathematical creativity in attempts to approach the limit, explaining the drive to complexity in living systems (Shadrin et al. 2013). The physical restraints (e.g. energy conservation) are thus the necessary prerequisites for forcing beings to explore the "Platonic mathematical world" (Penrose 2005), while the "Mental world" might arise out of necessity for memory and computing. Naturally the simple model captures only the general properties of genetic information processing as there are many features not included – epigenetics, rearrangements, roles of transposable and repetitive elements, recombination, multiple ploidy, etc.

In comparison with the other recently proposed explanation of Drake's rule (Sung et al. 2012), our model does not call for additional difficult-to-define entities like "molecular refinements", "drift barrier" or "effective population size" – the estimates of the later are admitted by the authors to be "fraught with difficulties". It is not clear how to simulate that evolutionary model in-silico, to perform its validation, because genome-wide functionality and conservation is not defined. Hence there is no specific model for selection actions, and there are many arbitrary parameters. However, a desirable feature of a "mechanistic" evolutionary model is the ability to simulate it, to evaluate its robustness in parameters space. Comparing Figure 1A of (Sung et al. 2012) with Figure 2 presented here, we can hypothesize that eukaryotes have lower *GI* density, in average, which is consistent with other observations (e.g. they exhibit weaker genomic conservation). Moreover, Figure 3 demonstrates that it can be advantageous to utilize the lower density *GI* storage. The *GI* storage strategy can be affected by particular demands for optimization: e.g. viruses or bacteria might prefer compact genomes with high *GI*, for faster replication or smaller physical size, utilizing the double stranded and overlapping coding and avoiding weakly conserved regulatory non-coding DNA.

The important consequence of our reasoning is that molecular evolution in average is not about a continuous increase of total *GI*. This suggests an explanation to a naive, but still valid question of why we see "living fossils" or don't see contemporary monkeys evolving into humans continuously, (anthropocentrically) assuming the latter have higher *GI*, while on the other hand we can observe an amazing morphological plasticity (e.g. dogs pedigrees or Cetacean evolution). Despite being "adaptive", for a given change of environment (selection demands), the evolution is not "progressive" in terms of total *GI*, since we posit that each species already have the maximum *GI*, allowed by the mutation rate, which is assumed to vary slowly. That also calls to revisit the popular evolutionary concept that genes are near their best functional performance – the performance is as good as allowed by the corresponding channel capacity – balancing at the brink of "chaos and order", so that a random mutation has high chances of being positive, in general. The dependence of the "evolvability" on the population size is also practically a "dogma" in traditional theories, which, might be a consequence of the unconstrained ("open-ended") opportunistic "Brownian" views on evolution. However, if a population is at the *GI* limit, so that an advance in one function must be associated with the "costs" to others, the role of the population size might be diminished, at least, as we showed, for the maintenance mode. In this scenario when an individual receives an advantageous mutation, its progeny will tolerate and keep more disadvantageous new mutations-hitchhikers (and outcomes of recombinations), which eventually will nullify the effect of the initial mutation. Qualitatively similar information "jamming" was also explored in the chapter "Conflict Resolution" in (Forsdyke 2011). It seems that strong dependencies on population size in traditional models lead to some contradictions with observations, such as Lewontin's "Paradox of Variation" (Lewontin 1974), not to mention the general trend that more evolved forms have smaller population sizes, in average. Ironically, we can draw the opposite scenario for evolvability vs. population size: without immediate negative effects, random mutations will degrade the rate in average. Hence the large population size in a long run can lead to the accumulation of variants, which increase the average mutation rate, leading to a degradation; then the way out is through the bottlenecks: the population must be regularly refreshed by the founding of subpopulations with decreased (below the average) rates. Such subpopulations will quickly gain an advantage and overcome the main population. In a sense it is the population genetic "ageing", analogously to a somatic ageing. In that case, the reproductive barriers, bottlenecks and speciation events are the necessities of evolution, required for the renewal and progress, rather than peculiar accidental features. We would like to remind that in this model positive mutations are abundant, so there is no need in a large population size or waiting time.

The adaptation to new selection demands then happens at the price of decrease of adaptation to other demands, the phenomenon well known to breeders (who now may attempt to select for the lower mutation rates also). For our model this can be imagined as a reshaping of genomic *GI* profile (and correspondingly a phenotype) while keeping the total *GI* constant. In biological interpretation it is the directional decrease of variability (reflected in the increase of corresponding *GI*) of one phenotypic feature (which is in demand), while increasing variability ("loosening up") of others. The traditional relative fitness function alone is unable to distinguish between such "reshaping" and "progressive" evolution modes, because the channel capacity notion is absent in traditional models (except for the somewhat analogous "error threshold" considerations, which are presumed to be narrowly applicable in some special cases). The general properties of such "reshaping selection" can be easily modeled with the suggested IT framework, to evaluate its basic features and the influences of diverse evolutionary strategies. In the case of eukaryotes we can expect that such evolutionary plasticity is residing mostly in non-coding regions with low *GI* density, since the fraction of beneficial mutations among random mutations is higher in weakly conserved regions.

# References

Chaitin G (2012) Proving Darwin: Making Biology Mathematical. Pantheon Books , New York.

Crow JF (1986) Basic concepts in population, quantitative, and evolutionary genetics. W.H. Freeman, New York, p. 273.

Drake JW (1991) A constant rate of spontaneous mutation in DNA based microbes. Proc Natl Acad Sci USA 88:7160–7164.

Drake JW, Charlesworth B, Charlesworth D, Crow JF (1998) Rates of Spontaneous Mutation. Genetics 148: 1667–1686.

Durrett R (2008) Probability Models for DNA Sequence Evolution. Springer, New York.

Feverati G, Musso F (2008) Evolutionary model with Turing machines. Phys Rev E Stat Nonlin Soft Matter Phys 77(6 Pt 1):061901.

Forsdyke DR (2011) Evolutionary bioinformatics. Second edition. Springer, New York Dordrecht Heidelberg London.

Haldane JBS (1957) The Cost of Natural Selection. J. Genet. 55:511-524.

Hardy GH (2003) Mendelian proportions in a mixed population. 1908. Yale J Biol Med 76: 79-80.

Hertz GZ, Stormo GD (1999) Identifying DNA and protein patterns with statistically significant alignments of multiple sequences. Bioinformatics 15(7-8):563-77.

Kimura M (1983) The neutral theory of molecular evolution. Cambridge University Press, New York.

Lewontin RC (1974) The genetic basis of evolutionary change. Columbia University Press, New York and London.

Moran PAP (1962) The statistical processes of evolutionary theory. Clarendon Press, Oxford.

Nowak MA (1992) What is a quasispecies? Trends Ecol Evol 7: 118-121.

Ohno, S (1970) Evolution by Gene Duplication. Springer-Verlag, New York.

Ohta T (1973) Slightly deleterious mutant substitutions in evolution. Nature 246(5428):96-8.

Ohta T, Gillespie JH (1996) Development of Neutral and Nearly Neutral Theories. Theor Popul Biol 49(2):128-42.

Penrose R (2005) The Road to Reality: A Complete Guide to the Laws of the Universe. Alfred A. Knopf, New York.

Schneider TD, Stephens RM (1990) Sequence Logos: a New Way to Display Consensus Sequences. Nucleic Acids Res 18:6097-6100.

Schneider TD, Stormo GD, Gold L, Ehrenfeucht A (1986) Information content of binding sites on nucleotide sequences. J Mol Biol 188:415-431.

Shadrin AA, Grigoriev A, Parkhomchuk DV (2013) Positional information storage in sequence patterns. Comput Mol Biosci. 3(2):18-26.

Shannon CE (1948) A Mathematical Theory of Communication. Bell System Technical Journal 27:379-423, 623-656.

Sung W, Ackerman MS, Miller SF, Doak TG, Lynch M (2012) Drift-barrier hypothesis and mutation-rate evolution. Proc Natl Acad Sci USA 109(45):18488-92.

Tajima F (1989) Statistical method for testing the neutral mutation hypothesis by DNA polymorphism. Genetics 123 (3): 585–95.